\documentstyle[12pt]{article}
\begin{document}
\hoffset = -1truecm
\voffset = -2truecm
\title{\bf  Ten arguments for the essence of special relativity}
\author{Guang-jiong Ni\thanks{Email:gj$ \_$ni$@$Yahoo.com}\\
\it \footnotesize \it Department of Physics, Fudan University,
Shanghai 200433, China\\
\it \footnotesize \it Department of Physics, Portland State
University, Portland, OR 97207, U.S.A.\thanks{mailing address}}
\date{}

\maketitle


\begin{abstract}
  In searching for the essence of special relativity, we have been
gradually accumulating ten arguments focusing on one fundamental
postulate based on quantum mechanics.A particle is always not pure.
It always contain two contradictory fields, $\varphi(\vec{x},t)$ and $\chi(\vec{x},t)$,which
are coupled together with the symmetry $\varphi(-\vec{x},-t)\longrightarrow\chi(\vec{x},t)$ and
$\chi(-\vec{x},-t)\longrightarrow\varphi(\vec{x},t)$.In a particle state $\varphi$ dominates $\chi$ as $|\varphi|>|\chi|$. But the enhancement of hiding $\chi$ ingredient in accompanying with the increase of particle velocity precisely accounts for the various
strange effects of special relativity. After newly defined space-time
inversion($\vec{x}\longrightarrow -\vec{x},t\longrightarrow -t$),
$\varphi(\vec{x},t)\longrightarrow\varphi(-\vec{x},-t)=\chi_{c}(\vec{x},t),
\chi(\vec{x},t)\longrightarrow\chi(-\vec{x},-t)=
\varphi_{c}(\vec{x},t),(|\chi_{c}|>|\varphi_{c}|)$,the particle transforms into its
antiparticle with the same momentum and (positive) energy.The above
symmetry should be regarded as a starting point to construct the
theory of special relativity, the relativistic quantum mechanics,the
quantum field theory and the particle physics,including the very
interesting superluminal theory for neutrino.[1]
\end{abstract}

\section{Motivation and belief}

  It was not until 1959 did physicists realize that the visual
image of so-called Lorentz contraction is not simply a contraction
along the motin direction. The misunderstanding lasting for 54 years
should be regarded as a lesson that the deduction method could be far
overwhelming the induction method and concrete analysis. However, we
should also raise a further question :\\
a. Why there is a Lorentz contraction?

  Besides the Lorentz contraction, various strange effects in special
relativity(SR) have been exhibiting themselves as mysteries of nature
for nearly 100 years. They are:\\
b.Why there is a limit $c=3\times 10^{8}m/s$ (speed of light in vacuum) for the
  velocity $v$ of any particle?(see, however, [25,26])\\
c.Why the inertial mass m of any particle increases with its velocity
  v and without a limit $[m=m_{0}{(1-\frac{v^{2}} {c^{2}})}^{-\frac{1} {2}}]$?\\
d.Why a moving clock accompanying the particle runs slower and slower
  when the particle velocity increases[clock frequency
$f=f_{0}{(1-\frac{v^{2}} {c^{2}})}^{\frac{1} {2}}$]?\\
  We should not be satisfied with the existing derivation of the above
four SR effects. Rather, we have been insisting on searching for a
deeper explanation for these effects,i.e.,for the essence of SR.
  Since 1905 till the recent years, there were many scholars trying
to derive SR by merely one relativistic postulate---the principle of
relativity (A) and abandoning the other relativistic postulate---
postulate of constancy of the speed of light(B). Eventually, all of
them failed. They did not realize that in 1905 Einstein had to
establish the kinematics of SR before the dynamics. So Eienstein
proposed B for establishing the Lorentz transformation ,the latter
at that time was no more than the mutual definition between the
coordinates of two inertial frames. If without B,
it would be meaningless to talk about A. Hence the meaning and value
of c must be fixed in advance by B.

  We should learn from the failure of many attempts and extract the
second lesson that ``A scientist should go beyond the `mere think',
one should raise the `intelligent questions' via experiments"( by
Galileo,quoted from [2]).

  Einstein did the best work in 20th century.But we may go further
beyond him because of the enormous accumulation of new experiments
since 1905.

  As is well known, the combination of SR with quantum mechanics
(QM) leads to relativistic quantum mechanics (RQM),quantum field
theory(QFT) and particle physics(PP) successfully. could it be a
consequence of ``complementarity" of two kinds of theory with
different essence (SR being clasical whereas QM being quantum)?

  No, let's look at the fact in biology. The combination of living
beings of different species can not reproduce their descendants.
It would be interesting to mention an exotic example. The mating
of a horse and a donkey gives birth of a mule,but the latter can
no longer have descendant.

  Now we know that the genetic factor in inheritance is DNA. So our
problems becomes the following:

  What is the DNA of RQM, QFT and PP inherited from the SR and QM
respectively?

  Further evidence about the conformity between SR and QM came from
de'Broglie. How could he derive his famous relation $\lambda=\frac{h} {p }$?

  First, he assumed that Einstein's relation $E=h\nu$ does hold for the
electron. However,on the other hand, the relation
$E=m_{0}c^{2}{(1-\frac{v^{2}} {c^{2}})}^{-\frac{1} {2}}$ in
SR implies that $E$ increases with the increase of velocity $v$, so does
$\nu$,a serious contradiction to the prediction of SR that the
frequency of moving clock should decrease (question d in section 1).

  To find the way out of the paradox, de'Broglie assumed next
that $\nu$ is not the frequency of a clock moving with the particle but
the frequency of a wave accompanying the particle (The frequency of
wave is measured at a fixed point in space). So the velocity of the
particle ,$v$, should be identified with the group velocity of wave:
\begin{equation}
v_g=\frac{d\omega} {dk}=v
\end{equation}

  Then after combination with the Lorentz transformation, the
de'Broglie relation
\begin{equation}
\lambda=\frac{h} {p},\qquad or \qquad p=\hbar k
\end{equation}
was derived.

  In short, de'Broglie derived the ``half" of quantum theory $p=\hbar k$
from another ``half" of quantum theory, $E=\hbar\omega$,in combination with
the whole theory of SR. His thinking clearly showed that the SR and
QM do have the common essence.

  The above two enlightenments are focusing on one belief that we
should strive to find the essence of SR on the basis of QM. The
problem is : where is the breakthrough point?

\section{The investigations on C,P,T problems}

  There are three discrete transformations discussed in QM:
\subsection{Space inversion (P)}
  The sign change of space coordinates ($\vec{x}\longrightarrow -\vec{x}$) in the wavefunction
(WF) of QM may lead to two eigenstates:
\begin{equation}
\psi_{\pm}(\vec{x},t)\longrightarrow \psi_{\pm}(-\vec{x},t)=\pm\psi_{\pm}(\vec{x},t)
\end{equation}
with eigenvalues 1 or -1 being even or odd parity.
\subsection{Time reversal (T)}
  Actually, the so-called time reversal (T) implies the reversal of
motion, which ascribes to the demand of invariance of Schrodinger
equation as follows:
\begin{eqnarray}
i\hbar\frac{\partial} {\partial t}\psi(\vec{x},t)&=&H\psi(\vec{x},t)\\
i\hbar\frac{\partial} {\partial t}[\psi^{\ast}(\vec{x},-t)]&=&H[\psi^{\ast}(\vec{x},-t)]
\end{eqnarray}

  In other words,the T invariance implies a relation of equivalence:
\begin{equation}
\psi(\vec{x},t)\sim \psi^{\ast}(\vec{x},-t)
\end{equation}

\subsection{Charge conjugation transformation}
  Aimming at changing the sign of charge(q) for a particle, e.g.,
from $q=-e$ for an electron to $q=e$ for the positron. one realize the C
transform in QM by demanding
\begin{equation}
\vec{p}+e\vec{A}=-i\hbar\bigtriangledown+e\vec{A}\longrightarrow -i\hbar\bigtriangledown-e\vec{A}
\end{equation}
which implies a complex conjugation on the WF:
\begin{equation}
\psi(\vec{x},t)\longrightarrow\psi^{\ast}(\vec{x},t)
\end{equation}

\subsection{CPT combined transform}
  If one considers the product of C,P,and T transformations together,
the complex conjugation contained in the T and C transforms will
cancel each other, yielding
\begin{equation}
\psi(\vec{x},t)
\longrightarrow
\mbox{\rm CPT}
\psi(\vec{x},t)=
\psi_{\mbox{\rm CPT}}
(\vec{x},t)
\sim
\psi(-\vec{x},-t)
\end{equation}
In the right side, the WF should be understood to describe the positron
(the antiparticle of electron). But it differs from the original WF
only in the sign change of x and t. What does it mean?

\subsection{Parity violation, CP(or T) violation and CPT theorem}
   The historical discovery of parity violation in 1956([3,4]) reveals
that both P and C symmetries are violated. Since 1964, it is found
that CP symmetry is also violated whereas CPT theorem remains valid,
which in turn implies the violation of T reversal symmetry, as further
verified by recent experiments[5].

  Therefore, the relation between a particle $|a>$ and its antiparticle
$|\bar{a}>$ is not
$|\bar{a}>=C|a>$ but [6]
\begin{equation}
|\bar{a}>=\mbox{\rm CPT}|a>
\end{equation}
which means nothing but precisely the Eq. (9).

   For an electron in free motion, its WF reads:
\begin{equation}
\psi_{e^{-}}(\vec{x},t)\sim \exp[\frac{i} {\hbar} (\vec{p}\cdot\vec{x}-Et)]
\end{equation}
Then Eq.(9) [or(10)] gives the WF of a positron as:
\begin{equation}
\psi_{e^{+}}(\vec{x},t)\sim \exp[-\frac{i} {\hbar} (\vec{p}\cdot\vec{x}-Et)]
\end{equation}
with the momentum $\vec{p}$ and energy $E>0$ precisely as that in the particle
state (11).

  The relation between Eqs. (11) and (12) should be viewed as a new
symmetry: `` The (newly defined) space-time inversion
($\vec{x}\longrightarrow -\vec{x},t\longrightarrow -t$)
is equivalent to particle-antiparticle transformation".
  In other words, the CPT theorem has already exhibited itself as a
basic postulate. The transformation of a particle to its antiparticle
( $\cal C$ ) is not something which can be defined independently but a direct
consequence of (newly defined) space-time inversion $\cal P\cal T$
($\vec{x}\longrightarrow -\vec{x},t\longrightarrow -t$)[7]:
\begin{equation}
\cal P \cal T=\cal C
\end{equation}
\subsection{Three understandings in physics[8]}
  (a) The observableness of a physical quantity must be related to
some symmetry or conservation law. Once it fails to do so, it will
cease to be an observable. The same is true for a transformation in
physics. Therefore, though C,P,and T are all clearcut transformations
in mathematics, they cease to be meaningful in physics.

  (b) There is an important difference between a ``theorem" and a ``law".
The various quantities contained in a theorem must be defined clearly
and unambiguously in advance before the theorem can be proved. On the
other hand, a law can often (not always) accommodate a definition of
a physical quantity which can only be defined unambiguously after the
law is verified by experiments.

  As a comparison, though Newton's equation $F=ma$ could be derived from
Lagrange variational principle or Hamilton principle, it is a law
rather than a theorem. The definition of inertial mass m is given by
the equation $m={F\over a}$ which should be verified by experiments.

  (c) A concept in physics should be expressed in terms of mathematical
language rigorously. If some concept can only be described by ordinary
language, it would be likely incorrect or at least not a deep one.

  For instance, the concepts like ``The positron carries the opposite
charge to that of electron" or `` In the vacuum all the (infinite)
negative states of electron are filled. Once a `hole' is created in the
`sea' , it would correspond to a positron",etc. are all incorrect. By
contrast, the correct concept of charge number $Q$ ($=q/e$ as the
substitution of charge $q$) now finds its expression as reflected in the
sign of phase of WF in Eqs. (11) and (12) , $Q=-1$ and 1 respectively.

\section{Einstein-Podolsky-Rosen paradox and antiparticle}

  In 1935, Einstein, Podolsky and Rosen(EPR) in a paper titled ``Can
quantum mechanical description of physical reality be considered
complete?"[9] raised a very strange question about two spinless
particles. Then after the seminar work by Bohm[10] and Bell[11],
physicists have been turning their attention to the puzzle of
entangled state,e.g.,the nonlocally correlated two-photon quantum
state over long distance, which was verified again and again by
experiments, especially in recent years[12,13]. However,another
puzzle in the original version of EPR paper was overlooked by the
majority of physicists but reemphasized by Guan[14] as follows:

  Consider two spinless particle in one-dimensional space with
positions $x_{i} (i=1,2)$ and momentum operators
${\hat{p}}_{i}=-ih\frac{\partial} {\partial x_{i}}$.Then the
commutation relation
\begin{equation}
[x_{1}-x_{2},\hat{p}_{1}+\hat{p}_{2}]=0
\end{equation}
implies that they have a common eigenstate with eigenvalues
\begin{equation}
x_1-x_2=a =const.
\end{equation}
\begin{equation}
p_1+p_2= 0  ,   p_2=-p_1
\end{equation}
   How strange the state is !? Two particles are moving in the opposite
momentum directions while keeping their distance unchanged. Incredible!
As stressed in Ref.[14],`` no one can figure out how to realize it." Now
we propose the following answer[15].

  If the WF of particle 1 is written as Eq. (11):
\begin{equation}
\psi(x_{1},t)=\exp[\frac{i} {\hbar}(p_{1}x_{1}-E_{1}t)]
\end{equation}
with $p_{1}>0$ and $E_{1}>0$, then the particle 2 must be an antiparticle:
\begin{equation}
\psi_{c}(x_{2},t)=\exp[\frac{i} {\hbar}(p_{2}x_{2}-E_{2}t)]
=\exp[-\frac{i} {\hbar}(p_{1}x_{2}-E_{1}t)]
\end{equation}
with $p_{2}=-p_{1}<0,E_{2}=-E_{1}<0$. But as shown in Eq.(12), we should view the WF
of ``negative energy state" of particle directly as the WF of its
antiparticle with corresponding operators:
\begin{equation}
\hat{p}_{c}=i\hbar\bigtriangledown,\hat{E}_{c}=-i\hbar\frac{\partial} {\partial t}
\end{equation}
which are the counterparts of that for the particle:
\begin{equation}
\hat{p}=-i\hbar\bigtriangledown, \hat{E}=i\hbar\frac{\partial} {\partial t}
\end{equation}
So the observed momentum and energy of antiparticle in state (18) are
$p_{1}$ and $E_{1}$ respectively, precisely the same as that of the particle state
(17). Now every thing is reasonable.

  If instead of Eq.(14), we consider the commutation relation:
\begin{equation}
[ x_1+x_2,\hat{p}_1-\hat{p}_2] = 0
\end{equation}
Then the correct answer turns out to be a particle and its
antiparticle moving in the opposite directions with momentum
$p_{1}$ vs $p^{c}_{2}=-p_{1}$ and positions $x_{1}$ vs
$x_{2}=-x_{1}$. Such kind of experiments have been performed many
times, say, in the process of $e^{+}e^{-}$ pair creation by a
high-energy photon in the vicinity of heavy neucleus. Especially,
a recent experiment reveals an entangled state of $K^{0} -
\bar{K}^{0}$ system[16] and provide a beautiful realization of
relation (21).

\section{Klien paradox in Klien-Gordon equation and antiparticle}

  In 1929, Klien discovered his famous paradox in Dirac equation,
resorting to the concept of ``hole" for introducing the antiparticle[17].
 Now we will discuss this paradox in Klien-Gordon(KG) equation with
much more simplicity and clarity[15]. As first shown in Ref.[18], the
KG Eq.
\begin{equation}
\left[i\hbar\frac{\partial} {\partial
t}-V(\vec{x})\right]^{2}\psi=-{c}^{2}{\hbar}^{2}{\bigtriangledown}^{2}\psi+m^{2}c^{4}\psi
\end{equation}
can be recast into two coupled Schrodinger equations of $\varphi$ and$\chi$:
\begin{eqnarray}
(i\hbar\frac{\partial} {\partial t}-V)\varphi&=&m c^{2}\varphi-\frac{\hbar^{2}} {2m}{\bigtriangledown}^{2}(\varphi+\chi)\nonumber\\
(i\hbar\frac{\partial} {\partial t}-V)\chi&=&-m c^{2}\chi+\frac{\hbar^{2}} {2m}{\bigtriangledown}^{2}(\chi+\varphi)
\end{eqnarray}
$$
\varphi=\frac{1} {2}[(1-\frac{V} {m c^{2}})\psi+i\frac{\hbar} {m c^{2}}\dot{\psi}]
$$
\begin{equation}
\chi=\frac{1} {2} [(1+\frac{V} {m{c}^{2}})\psi-i\frac{\hbar} {m{c}^{2}} \dot{\psi}]
\end{equation}
Eq. (23) is invariant under the (newly dedined) space-time inversion
($\vec{x}\longrightarrow -\vec{x},t\longrightarrow -t$) and transformation:
\begin{equation}
\varphi(-\vec{x},-t)\longrightarrow\chi(\vec{x},t),\chi(-\vec{x},-t)\longrightarrow\varphi(\vec{x},t)
\end{equation}
\begin{equation}
V(-\vec{x},-t)\longrightarrow -V(\vec{x},t)
\end{equation}
Let us consider a potential barrier in one dimensional space:
\begin{equation}
  V(x)=\left\{
     \begin{array}{ll}
       0,&\qquad x<0\\
       V_{0},&\qquad x\geq 0
     \end{array}
\right.
\end{equation}
The incident wave, reflected wave and transmitted wave are ,
respectively:
\begin{eqnarray}
\psi_{i}&=&a \exp[\frac{i} {\hbar}(px-Et)],\qquad(x\leq0 )\\
\psi_{r}&=&b \exp[\frac{i} {\hbar}(-px-Et)],\qquad(x\leq 0)\\
\psi_{t}&=&b' \exp[\frac{i} {\hbar}(p'x-Et)],\qquad(x\geq 0)
\end{eqnarray}
with $p>0,{p'}^{2}=\frac{{(E-V_{0})}^{2}} {c^{2}}-m^{2}c^{2}$. The continuation condition leads to
\begin{equation}
\frac{b} {a}=\frac{p-p'} {p+p'},\frac{b'} {a}=\frac{2p} {p+p'}
\end{equation}
The Klien paradox emerges when $V > E+mc$ .Since $p'=\pm\sqrt{\frac{{(V_{0}-E)}^{2}} {c^{2}}-m^{2}c^{2}}$ remains
real, the transmitting wave is oscillatory while the reflectivity of
incident wave reads:
\begin{eqnarray}
&&R={\left|\frac{b} {a}\right|}^{2}=\frac{{|p-p'|}^{2}} {{|p+p'|}^{2}}\nonumber\\
R&<&1,\mbox{\rm if}\qquad p'>0\\
R&>&1,\mbox{\rm if}\qquad p'<0
\end{eqnarray}
While the choice (32) is obviously unreasonable, the choice (33) seems
quite attractive.Then we have to realize why $p'<0$ and what happens in
this situation?

  For this purpose, we look back at Eqs. (22)-(24) and the accompanying
continuity equation[19]
\begin{equation}
\frac{\partial \rho} {\partial t}+\bigtriangledown\cdot\vec{j}=0
\end{equation}
\begin{equation}
\rho=\frac{i\hbar} {2mc^{2}}(\psi^{\ast}\dot{\psi}-\psi\dot{\psi}^{\ast})-\frac{V} {mc^{2}}\psi^{\ast}\psi=\varphi^{\ast}\varphi-\chi^{\ast}\chi
\end{equation}
\begin{eqnarray}
\vec{j}&=&\frac{i\hbar} {2m} (\psi \bigtriangledown\psi^{\ast}-\psi^{\ast}\bigtriangledown\psi)=\frac{i\hbar} {2m}
[(\varphi\bigtriangledown\varphi^{\ast}-\varphi^{\ast}\bigtriangledown\varphi)\nonumber\\
& &+(\chi\bigtriangledown\chi^{\ast}-\chi^{\ast}\bigtriangledown\chi)
+(\varphi\bigtriangledown\chi^{\ast}-\chi^{\ast}\bigtriangledown\varphi)
+(\chi\bigtriangledown\varphi^{\ast}-\varphi^{\ast}-\varphi^{\ast}\bigtriangledown\chi)]
\end{eqnarray}
Combining them with Eqs. (27)-(33), we find
\begin{eqnarray}
\rho_{i}&=&{|\varphi_{i}|}^{2}-{|\chi_{i}|}^{2}=\frac{E} {m c^{2}}{|a|}^{2}>0,j_{i}=\frac{p} {m}{|a|}^{2}>0\\
\rho_{r}&=&\frac{E} {m c^{2}}{|b|}^{2}>0,j_{r}=-\frac{\rho} {m}{|b|}^{2}<0\\
\rho_{t}&=&{|\varphi_{t}|}^{2}-{|\chi_{t}|}^{2}=\frac{(E-V_{0})} {m c^{2}}{|b'|}^{2}<0,j_{t}=\frac{p'} {m}{|b'|}^{2}
\end{eqnarray}
So we should demand $P'<0$ to get $j_{t}<0$ in conformity with $\rho_{t}<0$ and to
meet the requirement of Eqs.(14) ($j_{i}+j_{r}=j_{t}$) with  $|j_{r}|>j_{i}(|b|>|a|)$
and reflectivity $R>1$.

  The reason is clear. For an observer located at $x>0$, the energy of
particle in the transmitted wave should be measured with respect to the
local potential $V$ .So it has a negative energy $E'=E-V_{0}<0$ locally and
behaves as an antiparticle with its WF redefined as:
\begin{equation}
\psi_t\longrightarrow\psi^{c}_{t}=b'\exp[\frac{i} {\hbar}(p'x-E't)]=b'\exp[-\frac{i} {\hbar}(|p'|x-|E'|t)],\qquad (x>0)
\end{equation}
  According to Eq. (19), the energy and momentum of this antiparticle
are  $|E'|>0$ and $|p'|>0$ respectively. It moves to the right though $p'<0$
and $j_{t}<0$. Therefore, the KG Eq. is self-consistent even at one-particle
level to explain qualitatively the phenomenon of the $\pi^{+}\pi^{-}$pair creation
at strong field barrier bombarded by incident $\pi^{-}$ beam.

\section{The hidden antiparticle ingredient of a KG particle or Dirac particle}

\subsection{Klien-Gordon particle}

  Consider a freely moving wave packet for KG particle[19]:
\begin{eqnarray}
\psi(x,t)&=&{(4\sigma{\pi}^{3})}^{-\frac{1} {4}}\int^{\infty}_{-\infty} e^{-\frac{k^{2}} {2\sigma}}e^{i(kz-wt)}dk\nonumber\\
&\simeq&\frac{{(\frac{\sigma} {\pi})}^{\frac{1} {4}}} {{(1+\frac{i\sigma\hbar t} {m})}^{\frac{1} {2}}}
\exp\{\frac{-\sigma z^{2}} {2(1+\frac{i\sigma\hbar t} {m})}-i\frac{m c^{2} t} {\hbar}\}
\end{eqnarray}

Assuming$\frac{\sigma\hbar t} {m}\ll 1$to ignore the spreading of wave packet in low speed
($v<<c$) case,we perform a ``boost" transformation,i.e., to push the wave
packet to high speed($v\longrightarrow c$) case. Thus we see from the figures in [19]
that:

  (a) The width of packet shrinks-----Lorentz contraction;

  (b) The amplitude of $\rho$ increases-----``boost" effect;

  (c) The new observation is that both ${|\varphi|}^{2}$ and ${|\chi|}^{2}$ in $\rho$increase even
more sharply while keeping $|\varphi|>|\chi|$to preserve $|\varphi+\chi|=|\psi|$invariant.

   The ratio of hidden ingredient of $\chi$ to that of $\varphi$
is defined as
\begin{equation}
\left[R^{KG}_{free}\right]^2= \frac{\int^\infty_{-\infty}|\chi|^2
{\rm dz}} {\int_{-\infty}^\infty |\varphi|^2 {\rm dz}}
=\left[\frac{1-\sqrt{1-\frac{v^2} {c^2}}} {1+\sqrt{1-\frac{v^2}
{c^2}}}\right]^2 \stackrel{v\rightarrow c}{\longrightarrow}1
\end{equation}

  We are now in a position to answer the question (a) in section 1. It
is the enhancement of hidden $\chi$ field accompanying with the increase
of particle velocity v and the increasing coupling between $\chi$ and$\varphi$
fields lead to the upper bound for $v(v<c)$ and the appearance of Lorentz
contraction together with the boost effect.

  Next, consider a $\phi^{-}$ atom. The  $\pi^{-}$  meson in the Coulomb field of
neucleus has potential energy $V(r)=-\frac{z\alpha} {r}$and its $1S$ state energy being
a function of $Z$:
\begin{equation}
E^{KG}_{1s}=m c^{2}{\left[\frac{1} {2}+\sqrt{\frac{1} {4}-z^{2} {\alpha}^{2}}\right]}^{\frac{1} {2}}
\stackrel{z\rightarrow \frac{1} {2\alpha}}{\longrightarrow}\frac{1} {\sqrt{2}}m c^{2}
\end{equation}
while the relevant ratio reads:$(y=\sqrt{\frac{1} {4}-z^{2}{\alpha}^{2}})$
\begin{equation}
\left[R^{KG}_{1s}\right]^2=\frac{\int {|\chi|}^{2}d\vec{x}}
{\int{|\varphi|}^{2}d\vec{x}} =1-4{\left[2+{\left(y+\frac{1}
{2}\right)}^{\frac{1} {2}}+\frac{{(y+\frac{1} {2})}^{\frac{3}
{2}}} {2y}\right]}^{-1}
\end{equation}
which increases from 0 (when $Z=0$) to the upper limit 1 (when $Z\longrightarrow \frac{1} {2\alpha}$).

\subsection{Dirac particle}

  The WF of Dirac Eq. is a spinor with four components:
\begin{equation}
\psi=\left(
\begin{array}{lcr}
\varphi\\
\chi
\end{array}  \right)
\end{equation}
Here $\varphi$ and$\chi$ (each with two components),usually called as the ``positive"
and ``negative" energy components in the literature, are just the
counterparts of $\varphi$ and $\chi$ for KG particle.

  However, instead of Eq.(35), now we have
\begin{equation}
\rho_{\tiny{Dirac}}=\psi^\dag\psi=
\varphi^\dag\varphi+\chi^\dag\chi
\end{equation}
On the other hands, the invariant quantity during the boosting process is
\begin{equation}
\bar{\psi} \psi=\varphi^\dag \varphi- \chi^\dag \chi > 0
\end{equation}
Now both $\varphi^\dag\varphi$ and $\chi^\dag\chi$ are constrained under the $\rho$. In the limiting case,
$\varphi^\dag\varphi\geq\chi^\dag\chi\longrightarrow\frac{1} {2}\rho$. The relevant ratio for freely moving wave packet read:
\begin{equation}
\left[R^{Dirac}_{free}\right]^2=\frac{1-\sqrt{1-\frac{v^{2}}
{c^{2}}}} {1+\sqrt{1-\frac{v^{2}} {c^{2}}}}
\end{equation}
And the $1S$ state energy of Hydrogenlike atom is
\begin{equation}
E^{Dirac}_{1s}=m c^{2}\sqrt{1-z^{2} {\alpha}^{2}}\stackrel{z\rightarrow \frac{1} {\alpha}}{\longrightarrow}0
\end{equation}
with relevant ratio being
\begin{equation}
\left[R^{Dirac}_{1s}\right]^2=\frac{1-\sqrt{1-z^{2}{\alpha}^{2}}}
{1+\sqrt{1-z^{2}{\alpha}^{2}}}
\end{equation}

\subsection{Comparison between nonrelativistic QM and relativistic QM}

  Let us look at the coupling form of KG Eq.(23) and ignore all the $\chi$
field terms. Then we go back to the Schrodinger Eq. i.e., to the
nonrelativistic QM (NRQM).

  Now we are able to define the NRQM being the QM for particle with
all the hidden antiparticle ingredient ignored. In this case, neither
upper bound for particle velocity v nor lower bound for its energy E
in an external field exists. Meanwhile, the mass of particle, m,is an
invariant and there is no any relationship between $m$ and $E$.

  The situation changes radically in relativistic QM(RQM),as we just
see from either KG Eq. or Dirac Eq.. Once the hidden antiparticle
ingredient is taken into account,there must be an upper bound for the
velocity of particle, $v<c$, also a lower bound for its energy in an
external field, $E_{min}\geq 0$. Both these bounds are determined by the ratio
(R) of hidden antiparticle ingredient to that of particle ingredient:
$R\longrightarrow 1$. Meanwhile, the particle mass m becomes a variable and is
precisely proportional to the energy E of particle: $E=m c^{2}$.

  Now we begin to understand the essence of mass generation. It is also
the Essence of SR and could be ascribed to the equal existence of
particle with its antiparticle and the underlying symmetry shown as
Eq. (25).

\section{Derivation of special relativity}

  We are now in a position to derive the special relativity(SR) from
new point of view. Actually, what we shall do is nothing but looking
at the problem upside down[7].

  We will work at one inertial frame,i.e., the laboratory coordinate
system and begin with the NRQM,i.e., the Schrodinger equation. Then
comes the crucial step. We regard the basic symmetry Eq.(25) as the
only ``relativistic postulate" and inject it into the Schrodinger Eq.
for establishing Eq.(23),i.e., the KG Eq.. Let us stress the main
points as follows:

(a) A particle is always not pure. It always contains two contradictory
fields, $\varphi(x,t)$ and $\chi(x,t)$, which are coupled  eachother.

(b) Under the (newly defined) space-time inversion $(x\longrightarrow-x,t\longrightarrow-t)$,
\begin{equation}
\varphi(-\vec{x},-t)\longrightarrow\chi(\vec{x},t),\chi(-\vec{x},-t)\longrightarrow\varphi(\vec{x},t)
\end{equation}
together with the transformation property of external field:
\begin{equation}
V(-\vec{x},-t)\longrightarrow -V(\vec{x},t)
\end{equation}
the theory(equation) must be invariant.

(c) The new symmetry,i.e., the ``invariance under space-time inversion"
    exhibits itself as the only ``relativistic postulate".It should be
    assuned as the general feature of all relativistic theory.

(d) Of course, for dealing with the specific property of particle,we
    have to add another postulate of ``nonrelativistic" nature.For
    example, starting from Schrodinger Eq., we already assume the
    kinetic energy of spinless particle being in the form of $\frac{{\rho}^{2}} {2m}$and
    add a ``rest energy" term $m_{0}c^{2}_{1}$.On the other hand, we have to assume
    that the kinetic energy of spin 1/2 particle is in the form of $c_{2},\vec{\sigma},\vec{p}$
    while rest energy being $m_{0}c^{2}_{2}$. Based on the ansatz
    $\psi={\varphi \choose \chi}$ with $\varphi$or$\chi$being two component spinor, Dirac     equation can be derived
    similarly by the invariance postulate under the space-time inversion.

(e) After establishing the KG Eq.or Dirac Eq.,we obtain from the plane
    wave solution the following relation easily:
\begin{equation}
E^{2}=p^{2}c^{2}_{i}+m^{2}_{0}c^{4}_{i}, (i=1,2)
\end{equation}
    Then it is easy to see that the group velocity $V_{g}$of wave equals to
    the particle velocity $v$:
\begin{equation}
v_{g}=\frac{d\omega} {dk}=\frac{dE} {dp}=\frac{p {c}^{2}_{i}} {E}=v\stackrel{E\rightarrow\infty}{\longrightarrow}c_{i}
\end{equation}

    The experiments show that the limiting velocity of particle,$c_{i}$, is
    equal to the speed of light $c$.(Otherwise we would have no idea
    about space and time). Hence we have found relativistic dynamical
    law including Eqs.(53),(54) and
\begin{equation}
p=mv,   E=m c^{2},   m=\frac{m_{0}} {\sqrt{1-\frac{v^{2}} {c^{2}}}}
\end{equation}

(f)Next, we turn to relativistic kinematics. The main task is to find
   the relationship between the coordinates of two inertial frames,i.e.
   the Lorentz transformation. For this purpose, we need some
   invariance.Instead of the invariance of speed of light, this time we
   will resort to an invariance of nonrelativistic nature,i.e., the
   invariance of phase with respect to the coordinate transformation,
   (which was introduced by de'Broglie as a ``law of phase harmony"). A
   particle is moving in the laboratory system with velocity v,moment p
   along $x$ axis and energy $E$. We take a motion system resting on the
   particle and compare the phase of plane wave described in two
   systems:
\begin{equation}
\exp[\frac{i} {\hbar}(px-Et)]=\exp[\frac{i} {\hbar}(p'x'-E't')]=\exp(-\frac{i} {\hbar}m_{0}c^{2}t')
\end{equation}
   $(p'=0, E'=E_{0}=m_{0}c^{2})$.Substituting Eq.(55) into (56), we find
\begin{equation}
t'=\frac{t-\frac{vx} {c^{2}}} {\sqrt{1-\frac{v^{2}} {c^{2}}}}
\end{equation}
which is precisely the key relation in Lorentz transformation.

(g)In summary,since we adopt the approach from dynamics to kinematics,
   we can say more than Einstein in 1905. For instance, the two fields
   $\varphi(x,t)$ and $\chi(x,t)$ have opposite tendency of space-time evolution in
   their phases essentially:
\begin{eqnarray}
\varphi(\vec{x},t)&\sim& \exp[\frac{i} {\hbar}(\vec{p}\cdot\vec{x}-Et)]\\
\chi(\vec{x},t)&\sim&\exp[-\frac{i} {\hbar}(\vec{p}\cdot\vec{x}-Et)]
\end{eqnarray}
  So in a concrete particle state with $|\varphi|>|\chi|$, though $\chi$ has to obey
  $\varphi$ and follow the evolution as (58), it does exhibit itself as a drag
  and enhance the inertial mass.
    It is interesting to explain the time dilation effect in SR. According to Eqs.
  (58) and (59), in some sense, the time reading of the ``clock" for $\varphi$
  field is ``clockwise" whereas that for $\chi$ field is ``anticlockwise"
  essentially. Thus in accompanying with the increase of particle
  velocity, though the time reading remains ``clockwise", it runs slower
  and slower due to the enhancement of hidden $\chi$ field.

\section{Relativistic two-body stationary Schrodinger equation and \\ quarkonium}

   In particle physics,the heavy quarkonium $Q\bar{Q} (c\bar{c}\ or\ b\bar{b})$ is often
treated by Schrodinger equation with potential $V(r)=\sigma r$,where $r$ is the
distance between $Q$ and $\bar{Q}$ while $\sigma$ is called ``string tension constant".
Why can this ``nonrelativistic" potential model be so successful? Could
it be further improved at the level of QM in a simple manner?[20]

  We assume that the two-particle system should also be described by
two fields $\varphi(\vec{r}_{1},\vec{r}_{2},t)$ and $\chi(\vec{r}_{1},\vec{r}_{2},t)$
in coupling$(M=m_{1}+m_{2},\vec{r}=\vec{r}_{2}-\vec{r}_{1})$:
\begin{equation}
\left\{
   \begin{array}{lll}
    i\hbar\frac{\partial} {\partial t}\varphi&=&M\varphi-(\frac{\hbar^{2}} {2 m_{1}}\bigtriangledown^{2}_{1}+
          \frac{\hbar^{2}} {2 m_{2}}\bigtriangledown^{2}_{2})(\varphi+\chi)+V(r)(\varphi+\chi)\\
    i\hbar\frac{\partial} {\partial t}\chi&=&-M\chi+(\frac{\hbar^{2}} {2 m_{1}}\bigtriangledown^{2}_{1}+
          \frac{\hbar^{2}} {2 m_{2}}\bigtriangledown^{2}_{2})(\chi+\varphi)-V(r)(\chi+\varphi)
   \end{array}
\right.
\end{equation}
Eq.(60) is invariant under the space-time inversion with
\begin{equation}
\varphi(-\vec{r}_{1},-\vec{r}_{2},-t)\longrightarrow\chi(\vec{r}_{1},\vec{r}_{2},t),
\chi(-\vec{r}_{1},-\vec{r}_{2},-t)\longrightarrow\varphi(\vec{r},\vec{r},t)
\end{equation}
\begin{equation}
V(-\vec{r}_{1},-\vec{r}_{2},-t)\longrightarrow V(\vec{r}_{1},\vec{r}_{2},t)
\end{equation}
Note that, however, Eq.(62) is different from Eq.(26) where there is a
sign change in $V(\vec{x},t)$ under space-time inversion. This is because in
one-body equation $V$ is treated as an external field with the neucleus
as an inert core whereas here both two particles are turning into their
antiparticles under space-time inversion.

  Introducing the center-of-mass coordinate $\vec{R}=(m_{1}\vec{r}_{1}+m_{2}\vec{r}_{2})/M$ , reduced mass $\mu=m_{1} m_{2} /M$ and setting
\begin{equation}
\varphi=\Phi+\frac{i\hbar} {M c^{2}}\dot{\Phi},\chi=\Phi-\frac{i\hbar} {M c^{2}}\dot{\Phi}
\end{equation}
\begin{equation}
\Phi(\vec{R},\vec{r},t)=\psi(\vec{r})\exp[\frac{i} {\hbar}(\vec{P}\cdot\vec{R}-Et)]
\end{equation}
($\vec{P}$ is the momentum of center-of-mass and $E$ is the total energy of
 system) we arrive at
\begin{equation}
\left[-\frac{\hbar^{2}} {2\mu}{\bigtriangledown}^{2}_{\vec{r}}+V(\vec{r})\right]\psi(\vec{r})=\epsilon\psi(\vec{r})
\end{equation}
\begin{equation}
\epsilon=\frac{1} {2M c^{2}}(E^{2}-M^{2}c^{4}-P^{2}c^{2})
\end{equation}

We will set $P=0$ and denote the binding energy of system as $B=M c^{2} -E$.
Hence
\begin{equation}
B=M c^{2}\left[1-{(1+\frac{2\epsilon} {M c^{2}})}^{\frac{1} {2}}\right]
\end{equation}

  To our surprise,after taking the antiparticle effect into account,
the form of stationary Schrodinger Eq. undergoes no change but the
eigenvalue $\epsilon$ in the right side should not be directly identified with
$(-B)$ in the bound state. Rather, we should evaluate the $B$ fom $\epsilon$ via
Eq.(67).Note that there is a lower bound for $\epsilon : \epsilon_{min}=-M c^{2}/2$,i.e.,$E_{min}=0$.

  Let us use Eqs.(65),(66) for heavy quarkonium $Q\bar{Q}$ system. The
eigenvalues for S states reads:
\begin{equation}
\epsilon_{n}=\lambda_{n}{\left(\frac{\sigma^{2}} {2\mu}\right)}^{\frac{1} {3}},(n=1,2...)
\end{equation}
with $\lambda_{n}$ being the zero point of Airy function. So the total energy $E$ of
$Q\bar{Q}$ system should be calculated from (66), yielding
\begin{equation}
E_{n}=4\mu{\left[1+\frac{1} {2}\lambda_{n}{\left(\frac{\sigma^{2}} {2 \mu^{4}}\right)}^{\frac{1} {3}}\right]}^{\frac{1} {2}}
\end{equation}
($\mu=m/2=M/4$,$m$ being the quark mass).As a comparison,if one treated
 $\epsilon=E'-Mc$ directly, then the energy of $Q\bar{Q}$ sysyem would be
\begin{equation}
E'_{n}=4\mu'+\lambda_{n}{\left(\frac{{\sigma'}^{2}} {2\mu'}\right)}^{\frac{1} {3}}
\end{equation}
  The following Table 7.1 gives the comparison between the experimental
values $E^{exp}_{n}$ for six S states in Upsilon($b\bar{b}$) system and the theoretical
fitted values either from Eq.(69) ($E_{n}$) or from (70) ($E'_{n}$).

   \centerline{\bf Table 7.1 The energy levels of $S$ states in Upsilon($b\bar{b}$) system}
$$
\begin{tabular}{|c|c|c|c|c|c|c|}\hline
     n       &   1     &    2   &    3     &    4    &   5      &  6\\\hline
  E  (GeV)   & 9.46037 & 10.023 &  10.355  &  10.580 &   10.865 &  11.019\\ \hline
  E  (GeV)   & 9.46037 & 10.023 &  10.461  &  10.834 &   11.163 &  11.462\\ \hline
  E' (GeV)   & 9.46037 & 10.023 &  10.483  &  10.890 &   11.262 &  11.609\\ \hline
  $\lambda_{n}$ &  2.338  &  4.088 &   5.521  &   6.787 &    7.944 &   9.023\\ \hline
\end{tabular}
$$
  In fitting procedure,we have fixed two parameters $\sigma$ and $\mu$ by two
experimental values for $n=1$ and 2. Thus we find from Eq.(69) that
\begin{equation}
\left\{
  \begin{array}{lll}
  m_{b}&=& 2\mu=4.326 GeV\\
  \sigma&=&0.4530 Ge V^{2}
  \end{array}
\right.
\end{equation}
while from Eq.(70) that
\begin{equation}
\left\{
   \begin{array}{lll}
      m'_{b}&=&2\mu'=4.354 GeV\\
      \sigma'&=&0.3804 Ge V^{2}
    \end{array}
\right.
\end{equation}
The similar fit for Charmonium $J/\psi$($c\bar{c}$) system yields:
\begin{equation}
\left\{
  \begin{array}{lll}
    m_{c}&=&1.031 GeV\\
    \sigma&=&0.4183 Ge V^{2}
  \end{array}
\right.
\end{equation}
or
\begin{equation}
\left\{
   \begin{array}{lll}
     m'_{c}&=&1.155 GeV\\
     \sigma'&=&0.2099 Ge V^{2}
   \end{array}
\right.
\end{equation}
  We can see that the value for $\sigma'$ given by (72) and (74) are in
discrepancy too big while that for $\sigma$ given by (71) and (73) seems
much better.

  In summary, now we understand two points:

(a) The reason why the potential model in ``nonrelativistic" stationaary
Schrodinger Eq. was so successful for describing the $Q\bar{Q}$ system lies in the fact that
the potential $V(r)$ is actually a ``four-dimensional scalar potential" satistying Eq.(62)
rather than a ``vector potential" satisfying Eq.(26).In recent years,
this kind of confining scalar potential $V(r)\sim \sigma r$ can be derived from
quantum chromodynamics(QCD) as discussed by Brambilla and Faustov[21,22].

(b) We are able to make ``relativistic correction" on ``nonrelativistic"
model at QM level by using Eq.(67),as shown at the improvement of
Eq.(69) vs (70),in a very simple way.

\section{Field operators in quantum field theory}

 The vector potential of classical electromagnetic field is real:
\begin{equation}
\vec{A}(\vec{x},t)=\frac{1} {\sqrt{V}}\sum\limits_{\vec{k},\lambda}{\vec{\varepsilon}}^{(\lambda)}_{\vec{k}}
[C_{\vec{k}\lambda}(t)e^{i\vec{k}\cdot\vec{x}}+C^{\ast}_{\vec{k}\lambda}(t)e^{-i\vec{k}\cdot\vec{x}}]
\end{equation}
which can be quantized into a Hermitian field operator in quantum field
theory(QFT) as follows:
\begin{equation}
\hat{\vec{A}}(\vec{x},t)=c\sqrt{\frac{\hbar} {2\omega V}}\sum\limits_{\vec{k},\lambda}
{\vec{\varepsilon}}^{(\lambda)}_{\vec{k}}[\hat{a_{\vec{k}\lambda}}(t)e^{i\vec{k}\cdot\vec{x}}+
{\hat{a}}^{\dag}_{\vec{k}\lambda}(t)e^{-i\vec{k}\cdot\vec{x}}]
\end{equation}
Here the amplitudes $C_{\vec{k}\lambda}(t)$ in (75) had been promoted into operators
$\hat{a}_{\vec{k}\lambda}(t)$ in Fock space on photon with commutation relations:
\begin{equation}
[{\hat{a}}_{\vec{k}\lambda}(t),{\hat{a}}^{\dag}_{\vec{k'}\lambda'}(t)]=\delta_{\vec{k}\vec{k'}}\delta_{\lambda\lambda'}
\end{equation}
  Next, for ``classical" Dirac field, one has
\begin{equation}
\psi(\vec{k},t)=\frac{1}
{\sqrt{V}}\sum\limits_{\vec{p}}\sqrt{\frac{m c^{2}} {|E|}}
\{\sum\limits_{r=1,2(E>0)}b^{(r)}_{\vec{p}}u^{(r)}(\vec{p})e^{\frac{i}
{\hbar}(\vec{p}\cdot\vec{x}-Et)}+
\sum\limits_{r=3,4(E<0)}b^{(r)}_{\vec{p}}u^{(r)}(\vec{p})e^{\frac{i}
{\hbar}(\vec{p}\cdot\vec{x}-Et)}\}
\end{equation}
Then, depending on the ``hole" concept,one defined the operators
for antiparticle as:
\begin{equation}
{\hat{d}}^{(s){\dag}}_{\vec{p}}=\mp{\hat{b}}^{(r)}_{-\vec{p}}
(s=1,r=4;s=2,r=3), \qquad v^{(s)}(\vec{p})=\mp u^{(r)}(-\vec{p})
\end{equation}
to get the field operator for Dirac particles:
\begin{equation}
\hat{\psi}(\vec{x},t)=\frac{1} {\sqrt{V}}
\sum\limits_{\vec{p}}\sum\limits_{s=1,2(E>0)}\sqrt{\frac{m c^{2}} {E}}
\{{\hat{b}}^{(s)}_{\vec{p}}U^{(s)}(\vec{p})e^{\frac{i} {\hbar}(\vec{p}\cdot\vec{x}-Et)}+
{\hat{d}}^{(s){\dag}}_{\vec{p}}v^{(s)}(\vec{p})e^{-\frac{i} {\hbar}(\vec{p}\cdot\vec{x}-Et)}\}
\end{equation}
and anticommutation relations:
\begin{equation}
[\hat{b}^{(s)}_{\vec{p}},\hat{b}^{\dag(s')}_{\vec{p'}}]_{+}
=[\hat{d}^{\dag(s)}_{\vec{p}},\hat{d}^{(s')}_{\vec{p'}}]_{+}
=\delta_{\vec{p}\vec{p'}}\delta_{\vec{s}\vec{s'}}
\end{equation}
  However, for complex KG field, though there are infinite solutions
with negative energy, the ``hole" concept can't work. One had to
quantize it similar to that in Eq.(80) by defining
\begin{equation}
\hat{\Phi}(\vec{x},t)=\sum\limits_{\vec{p}}\frac{1} {\sqrt{2\omega_{\vec{p}}V}}
[\hat{a}_{\vec{p}}e^{\frac{i} {\hbar}(\vec{p}\cdot\vec{x}-Et)}+{\hat{b}}^{\dag}_{\vec{p}}
e^{-\frac{i} {\hbar}(\vec{p}\cdot\vec{x}-Et)}]
\end{equation}
with
\begin{equation}
[\hat{a}_{\vec{p}},\hat{a}^{\dag}_{\vec{p'}}]=[\hat{b}_{\vec{p}},\hat{b}^{\dag}_{\vec{p'}}]=\delta_{\vec{p}\vec{p'}}
\end{equation}
to describe the annihilation and creation of KG particle and its
antiparticle.

  The question remains as ``what is the reason for doing so?" Or more
generally,what is the unified basis for the definition of above three
kind of field operators ($\hat{\vec{A}},\hat{\psi}$ and  $\hat{\phi}$)?

  We propose the following answer[7].It is just the (newly defined)
``invariance of field operator under space-time inversion" that ensures
its correctness. Say, for Eq.(8), it reads:
\begin{equation}
\hat{\phi}(-\vec{x},-t)=\hat{\phi}(\vec{x},t)
\end{equation}
which contains the transformation
\begin{equation}
\hat{a}_{\vec{p}}\rightleftharpoons\hat{b}^{\dag}_{\vec{p}},(\vec{x}\longrightarrow -\vec{x},
t\longrightarrow -t)
\end{equation}
as an intuitive complement to Eq.(13).

\section{Connection between spin and statistics}

  We should address a further question in the previous section:``Why one
must quantize the KG field by commutation relation whereas the Dirac
field by anticommutation relation?"

  As a preparation, we first note that Dirac equation has two forms[7].
In Pauli metric,it is usually written as
\begin{equation}
(\gamma_{\mu}\frac{\partial} {\partial x_{\mu}}+m)\psi(x)=0
\end{equation}
with $\psi\sim  \left( \begin{array}{c}
\varphi\\
\chi\end{array} \right) e^{-\frac{iEt} {\hbar}}.$ As discussed in
previous section,if $|\frac{\varphi} {\chi}|>1,E>0,\psi$ describes
the electron.If $|\frac{\chi} {\varphi}|>1,E<0$, $\psi$ describes
the positron.Let us perform a space-time inversion:
$\psi(x)\longrightarrow \psi(-x)=\psi_{c}(x)$,then Eq.(86) changes
to
\begin{equation}
(\gamma_{\mu}\frac{\partial} {\partial x_{\mu}}-m)\psi_{c}(x)=0
\end{equation}
which remains effective with $\psi_{c}\sim{\chi_{c} \choose
\varphi_{c}}e^{iEt \over \hbar}$. If $|\frac{\chi_{c}}
{\varphi_{c}}|>1, E>0$, $\psi_{c}$ describes the positron.If
$|\frac{\varphi_{c}} {\chi_{c}}|>1,E<0$, $\psi_{c}$ describes the
electron.The difference between Eqs.(86) and (87),or between
$\psi$ and $\psi_{c}$,is merely a representation
transformation:$\psi\longrightarrow\psi_{c}=\gamma_{5}\psi,
(\gamma_{5}=\gamma_{1}\gamma_{2}\gamma_{3}\gamma_{4}).$

  We are now in a position to prove the connection between spin and
statistics in QFT. Beginning from the ``Principle of microscopic
causality",one can first arrive at the general commutation relation for
KG field as
\begin{equation}
[\hat{\phi}(x),\hat{\phi}^\dag(y)]_{\omega}=i\hbar c \bigtriangleup(x-y)
\end{equation}
with
\begin{equation}
\bigtriangleup(x)=0,x^{2}>0
\end{equation}
which means that there is no connection between two points with space-
like distance. However,the subscript $\omega$ of the bracket in the left side
is not fixed yet either to be $\omega=-1$ (commutation relation) or $\omega=1$
(anticommutation relation).

  Now we use the ``invariance under the space-time inversion" to fix the
$\omega=-1$. Actually,note that $\bigtriangleup(-x)=-\bigtriangleup(x)$ and
\begin{equation}
\hat{\phi}(x){\hat{\phi}}^\dag(y)\stackrel {x\rightarrow -x} {\longrightarrow}
\hat{\phi}^\dag(-y)\hat{\phi}(-x)=\hat{\phi}^\dag(y)\hat{\phi}(x)
\end{equation}
( The order of operator product has to be reversed.)
 Then $\omega=-1$  follows immediately.

  Similarly, for Dirac field, one first arrives at
\begin{equation}
[\hat{\psi}(x),\hat{\bar{\psi}}(y)]_{\omega}=-i(\gamma_{\mu}\frac{\partial} {\partial x_{\mu}}
-\frac{mc} {\hbar})\bigtriangleup(x-y)
\end{equation}
Next, under the space-time inversion, we have
\begin{equation}
[\hat{\psi}_{c}(x),\hat{\bar{\psi}}_{c}(y)]_{\omega}=\left\{
\begin{array}{ll}
-i(\gamma_{\mu}\frac{\partial} {\partial x_{\mu}}+\frac{mc} {\hbar})\bigtriangleup(x-y),&{\rm if}\qquad \omega=+1\\
i(\gamma_{\mu}\frac{\partial} {\partial x_{\mu}}+\frac{mc} {\hbar})\bigtriangleup(x-y),&{\rm if}\qquad  \omega=-1
\end{array}
\right.
\end{equation}

Hence the invariance demands $\omega=1$ up to a representation transformation
($\psi\longrightarrow\psi_{c}$). the other choice $\omega=-1$ is certainly excluded.

  In summary, the combination of the principle of microscopic causality
with the invariance under the (newly defined)space-time inversion leads
to the correct connection between spin and statistics unambiguously.

\section{The Feynman propagator for electron}

  In the nonrelativistic QM (NRQM), the evolution of WF for a particle
is described by Feynman as
\begin{eqnarray}
\psi(x_{b},t_{b})&=&\int K(b,a)\psi(x_{a},t_{a})dx_{a}\\
K(b,a)&=&\sum\limits_{a\rightarrow b, (a\| paths)}const\cdot \exp(\frac{i} {\hbar} S[x(t)])
\end{eqnarray}
with classical action $S[x(t)]$ calculated along a path connecting points
$a$ and $b$. However, all paths, in spite of their arbitrariness, must go
forward in time.

  This can also be seen from the Green function of Schrodinger Eq.:
\begin{equation}
(i\hbar\frac{\partial} {\partial t}-\hat{H})G(x,t|x',t')=\delta(x-x')\delta(t-t')
\end{equation}
\begin{equation}
G(x,t|x',t')=-\frac{i} {\hbar} K(x,t|x',t')\theta(t-t')
\end{equation}
\begin{equation}
\theta(t-t')=\left\{
\begin{array}{ll}
1,& t>t'\\
0,& t<t'
\end{array}
\right.
\end{equation}
Hence the Green function $G$ given by (96) is precisely the same as the
Feynman kernel function $K$ given by (94), except the existence of $\theta$ function
showing explicitly that the reversal of evolution in time is not
allowed. We remind our readers onceagain of the misnomer of so-called
``time-reversal" in NRQM as discussed in section 2.

  The situation is radically different in relativistic QM (RQM) where
the Feynman propagator for electron, $K_{F}(x,x')$, is defined as
\begin{equation}
(\gamma_{\mu}\frac{\partial} {\partial x_{\mu}}+m)K_{F}(x,x')=-i\delta^{(4)}(x-x')
\end{equation}
\begin{equation}
K_{F}(x,x')=\sum\limits_{\vec{p},s}\frac{m} {EV} \{ u^{(s)}(\vec{p}){\bar{u}}^{(s)}(\vec{p})
e^{i p\cdot(x-x')}\theta(t-t')-v^{(s)}(\vec{p}){\bar{v}}^{(s)}(\vec{p})
e^{-ip\cdot(x-x')}\theta(t'-t)\}
\end{equation}
Let us perform a space-time inversion [7]:
\begin{eqnarray}
K_{F}(x,x')&\longrightarrow& K_{F}(-x,-x')=K^{c}_{F}(x,x')\nonumber\\
&=&\sum\limits_{\vec{p},s}\frac{m} {EV} \{
u^{(s)}(\vec{p}){\bar{u}}^{(s)}(\vec{p})e^{-ip\cdot(x-x')}\theta(t'-t)
-v^{(s)}(\vec{p})\bar{v}^{(s)}(\vec{p})e^{ip\cdot(x-x')}\theta(t-t')\}\nonumber\\
&=&\gamma_{5}K_{F}(x,x')\gamma_{5}\sim K_{F}(x,x')
\end{eqnarray}
which means that $K_{F}(x,x')$ is invariant under the space-time inversion
up to a representation transformation.

\section{Superluminal theory for neutrino}

  Einstein's theory of SR and the principle of causality imply that the
speed of any moving object cannot exceed that of light,$c$.However,there
were many attempts in literature discussing the particle moving with
speed $u>c$, called as superluminal particle or tachyon.In recent years,
the experimental data show that the measured neutrino mass-square is
negative.It was reported in Ref.[23] that
\begin{equation}
m^{2}(\nu_{e})=-2.5+3.3eV^{2},
m^{2}(\nu_{\mu})=-0.016 + 0.023 MeV^{2}
\end{equation}
which though far from accurate, does strongly hint that neutrino
might be a superluminal particle having energy and momentum
relation as
\begin{equation}
E^{2}=c^{2} p^{2} - m^{2}_{s} c^{4}
\end{equation}
with the ``proper mass" $m_{s}$ being real and positive.Then it is
easy to prove that
\begin{equation}
p = m_{s}u{\left( \frac{u^{2}} {c^{2}} -1 \right)}^{-\frac{1} {2}},
E = m_{s}c^{2}{\left(\frac{u^{2}} {c^{2}}-1\right)}^{-\frac{1} {2}}
\end{equation}
with $u$ being the velocity of tachyon.(see,e.g.,[24]).

  To derive the relation (102) from a quantum theory and following
Dirac's idea, Chang proposed a Dirac-type equation as follows[25]:
\begin{equation}
i\hbar\frac{\partial} {\partial t}\psi_{s}=[-c\vec{\alpha}\cdot\hat{\vec{p}}+
\beta_{s}m_{s}c^{2}]\psi_{s}
\end{equation}
$$
\alpha_{i}=\left(
\begin{array}{cc}
0 & \sigma_{i}\\
\sigma_{i} & 0
\end{array}
\right), \qquad
\beta_{s}=\left(
\begin{array}{cc}
0 & I\\
-I & 0
\end{array}
\right)$$
which gives the relation (102) straightforwardly.

  At first sight,the nonhermitian property of $\beta_{s}$ would obstruct Eq.(104)
from being accepted. However, after careful examination[26], it is
shown that Eq.(104) precisely satisfies the basic symmetry
discussed in previous sections.Indeed, setting $\psi_{s}={\varphi
\choose \chi}$, we can write Eq.(104) as
\begin{equation}
\left\{
\begin{array}{l}
i\hbar\frac{\partial} {\partial t}\varphi=ic\hbar\vec{\sigma}\cdot\bigtriangledown\chi+
m_{s}c^{2}\chi\\
i\hbar\frac{\partial} {\partial t}\chi=ic\hbar\vec{\sigma}\cdot\bigtriangledown\varphi-
m_{s}c^{2}\varphi
\end{array}
\right.
\end{equation}
Evidently, it is invariant under the space-time inversion with
transformation (25), just like Dirac Eq.,which reads:
\begin{equation}
\left\{
\begin{array}{l}
i\hbar\frac{\partial} {\partial t}\varphi_{D}=ic\hbar\vec{\sigma}\cdot\bigtriangledown\chi_{D}+
m_{0}c^{2}\varphi_{D}\\
i\hbar\frac{\partial} {\partial t}\chi_{D}=ic\hbar\vec{\sigma}\cdot\bigtriangledown\varphi_{D}-
m_{0}c^{2}\chi_{D}
\end{array}
\right.
\end{equation}
But what is the difference between them?
After introducing $\xi=\frac{1} {\sqrt{2}}(\varphi+\chi)$ and $\eta=\frac{1} {\sqrt{2}}(\varphi-\chi)$ to recast (105) and (106) into
\begin{equation}
\left\{
\begin{array}{l}
i\hbar\frac{\partial} {\partial t}\xi=ic\hbar\vec{\sigma}\cdot\bigtriangledown\xi-
m_{s}c^{2}\eta\\
i\hbar\frac{\partial} {\partial t}\eta=-ic\hbar\vec{\sigma}\cdot\bigtriangledown\eta+
m_{s}c^{2}\xi
\end{array}
\right.
\end{equation}
and
\begin{equation}
\left\{
\begin{array}{l}
i\hbar\frac{\partial} {\partial t}\xi_{D}=ic\hbar\vec{\sigma}\cdot\bigtriangledown\xi_{D}+
m_{0}c^{2}\eta_{D}\\
i\hbar\frac{\partial} {\partial t}\eta_{D}=-ic\hbar\vec{\sigma}\cdot\bigtriangledown\eta_{D}+
m_{0}c^{2}\xi_{D}
\end{array}
\right.
\end{equation}
respectively,we see that while Eq. (108) is invariant under the space
inversion ($\vec{x}\longrightarrow -\vec{x},t\longrightarrow t$) and transformation
\begin{equation}
\xi_{D}(-\vec{x},t)\longrightarrow\eta_{D}(\vec{x},t),
\eta_{D}(-\vec{x},t)\longrightarrow\xi_{D}(\vec{x},t)
\end{equation}
Eq.(107) fails to do so because of the mass terms with opposite signs
before them.Hence we realize that the violation of hermitian property
is due to the violation of parity which was discovered in 1956[3,4]. The
new observation is that the maximum parity violation is triggered by
nonzero mass ($m_{s}$) which in turn implies that neutrino must be a
superluminal particle with permanent helicity.

  In Ref.[26], after introducing two parameters $R=\sqrt{\frac{\chi^\dag\chi} {\varphi^\dag\varphi}}$
and $W=\sqrt{\frac{\eta^\dag\eta} {\xi^\dag\xi}}$,and analizing
the anticorrelation between them in the whole range of particle
speed ($0 < u < \infty$),we are able to understand the intrinsic
(dynamical) reason responsible for the strange kinematic behavior
of particle with $u<c$( i.e.,the SR effect as discussed in section
6) as well as that with $u>c$ as shown in Eq.(103). See also
[27-30].

\section{Summary and discussion}

(a) The special relativity (SR) and quantum mechanics (QM) are in
conformity in essence, so their combination can lead to the vigorous
relativistic QM (RQM),quantum field theory (QFT) and particle physics
(PP). The genes(DNA) in their inheritance are,respectively:

$$\begin{array}{llll} QM: & \hat{\vec p}=-i\hbar\bigtriangledown &
SR:  &  \hat{\vec p}_c=i\hbar\bigtriangledown \cr
  &  \hat{E}=i\hbar\frac{\partial} {\partial t} &
&
  \hat{E}_c=-i\hbar\frac{\partial} {\partial t}
\end{array}
$$
(b) A particle is always not pure. Its wave function(WF) always
contain two contradictory fields, $\varphi(x,t)$ and $\chi(x,t)$.
Essentially,
\begin{eqnarray}
\varphi&\sim&\exp[\frac{i} {\hbar}(\vec{p}\cdot\vec{x}-Et)],
\qquad(E>0)\nonumber\\
\chi&\sim&\exp[-\frac{i}
{\hbar}(\vec{p}\cdot\vec{x}-Et)],\qquad(E>0)\nonumber
\end{eqnarray}
If $|\varphi|>|\chi|$, the particle exhibits itself as a
``particle" with
$$\psi\sim \varphi\sim \chi\sim \exp[\frac{i} {\hbar}(\vec{p}\cdot\vec{x}-Et)],\qquad(E>0),$$
If $|\chi_{c}|>|\varphi_{c}$, the particle exhibits itself as an
``antiparticle" with
$$\psi_{c}\sim \chi_{c}\sim \varphi_{c}\sim \exp[-\frac{i} {\hbar}(\vec{p}\cdot\vec{x}-Et)],\qquad(E>0),$$
(c) There is no any ``negative energy state", no ``sea" or ``hole" at all.
The historical mission of ``hole" theory is coming to an end.\\
(d) We should not regard the $\vec{x}$ in the WF $\psi(x,t)$ as the coordinate of
``point particle" before the measurement. Rather, $\vec{x}$ and $t$ are the
flowing coordinates of fields [27,28].\\
(e) The CPT theorem already exhibits itself as a basic postulate:\\
$$\cal P \cal T=\cal C $$
The operator at the left side ($\cal P\cal T$) just means
$x\longrightarrow -x,t\longrightarrow -t$,the newly
defined space-time inversion.The operator at the right side ($\cal C$) means
the physical particle-antiparticle transformation, a definition being
contained right here (not coming from elsewhere).\\
(f) The above symmetry should be pushed forward to
\begin{eqnarray}
\varphi(-\vec{x},-t)&\longrightarrow&\chi(\vec{x},t)\nonumber\\
\chi(-\vec{x},-t)&\longrightarrow&\varphi(\vec{x},t)\nonumber
\end{eqnarray}
forming a starting point to construct the theory of SR,the RQM,the QFT
and the PP. The superluminal theory for neutrino is just the new and
interesting manifestation of the subtlety of QM and the basic symmetry.\\
(g) Actually,the basic symmetry lies in the essential equivalence
of ``$i$" and ``$-i$", which is relevant to the fundamental
interpretation of QM. Einstein had pointed out that ``space and
time are closely related and inseparable." We could add that
``space-time and matter are also closely related and inseparable".
It is time for the revival of ``Ether"[8,15,31,32,33].

\vskip.2in

\leftline{\bf References}
\begin{enumerate}
\item Author's note:\qquad In 21-23 June 2000, I attended the 23rd International
  Workshop on the Fundamental Problems of High Energy Physics and Field
  Theory held at Protvino, Russia and presented a lecture titled ``Nine
  arguments for the essence of special relativity".Fortunately,  in July
  2000, Dr.T.Chang brought a equation for neutrino as a superluminal
  particle to my attention,  which results in collaboration of us and
  two papers[25, 26]. So I add section 11 to this paper and change the
  title. This paper was published in Proceedings Edited by I. V.
  Filimonova and V. A. Petrov, pp. 275-292 and then slightly
  modified.
\item Wu, C.S., Lecture at Padova University, 1986.in ``Scientific career in
  half century--Collection of C.S.Wu and J.L.Yuan", ed.by D.Feng and
  D.Lu, (Press of Nanking University, 1992).
\item Lee, T.D.\& Yang, C.N.Phys.Rev.104, 254(1956).
\item Wu, C.S.et al. Phys.Rev.105, 1413 (1957).
\item Angelopoulos, A. et al (CPLEAR Collaboration), Phys.Lett.B444, 43;52
  (1998).
\item Lee, T.D.\& Wu, C.S.Ann.Rev.Nucl.Sci, 15, 381 (1965).
\item Ni, G.-j.\& Chen, S.-q. On the essence of special relativity, J.Fudan
  University(Natural Science) 35,  325 (1996); English version is in ``Photon
  and Poincare Group",  ed.by V.Dvoeglazov, (Nova Publisher, Inc.1999),
  Chapt.III.
\item Ni, G.-j. To enjoy the morning flower in the evening -- Is special
  relativity a classical theory? Kexue (Science), 50 (1) 29 (1998); -- Where
  is the subtlety of quantum mechanics? ibid,  50(2) 38 (1998); -- What does
  the infinity in physics imply? ibid,  50 (3) 36 (1998). English versions are
  in ``Photon: Old Problems in Light of New Ideas", sd.by V. Dvoeglazov
  (Nova Publisher, Inc. 2000), Chapt.III.
\item Einstein, A., Podolsky, B.\& Rosen, N., Phys. Rev. 47,  777
  (1935).
\item Bohm, D.``Quantum Theory"(Prentice Hall, 1951).
\item Bell, J.S.On Einstein-Podolsky-Rosen paradox, Physics (Long Island
   City,  N.Y.)1, 195 (1964).
\item Weihs, G. et al.Phys.Rev.Lett.81, 5039(1998).
\item Tittel, W. et al.Phys.Rev.A57, 3229(1998);Phys.Rev.Lett.81, 3563(1998).
\item Guan, H.``Basic Concept in Quantum Mechanics"(Higher Education Press,
   Beijing, 1990), Chapt.7.
\item Ni, G.-j., Guan, H., Zhou, W.-m.\& Yan, J.Antiparticle in the light of
   Einstein-Podolsky-Rosen paradox and Klein paradox,  Chin. Phys. Lett.
   17,   393-395(2000).
\item Apostolakis, A.et al.(CPLEAR Collaboration), Phys.Lett.B422, 339(1998).
\item Klein, O.Z.Phys.53, 157(1929).
\item Feshbach, H.\& Villars, F.Rev.Mod.Phys.30, 24(1958).
\item Ni, G.-j., Zhou, W.-m.\& Yan, J.Comparison among Klein-Gordon equation,
   Dirac equation and relativistic stationary Schrodinger equation,
   Proceeding of International Workshop,  Lorentz Group, CPT, and
   Neutrinos, Zacatecas,  Mexico,  June 23-26,  1999 (World Scienfic, 2000).
\item Ni, G.-j.\& Chen, S.-q.Relativistic stationary Schrodinger equation,
   J.Fudan University(Natural Science)36, 247(1997); Ni, G.-j. Relativistic
   stationary Schrodinger equation and its applications, hep-th/9708156,
   Proceeding of International Workshop, Lorentz Group, CPT, and
   Neutrinos,  Zacatecas,  Mexico,  June 23-26,  1999 (World Scientific, 2000).
\item Brambilla, N.pNRQCD:a new effctive theory for quarkonium, Lecture at
   the 23rd International Workshop on the Fundamental Problem of High
   Energy Physics and Field Theory, 22 June, 2000, Proceedings:
   pp106-126.
\item Ebert D., Faustov R. N. and Galkin V. O., ibid pp68-78.
\item Review of Particle Physics, Euro.Phys.Journ.C15, 350-353(2000).
\item Chodos A. et al., Phys. lett. B150, 431 (1985); Ciborowski J. and Rembielinski, J. Europ. Phys. J. V8, 157 (1999);
Chang, T. Beyond relativity,  Preprint, 2000.
\item Chang, T.\& Ni.G.-j, An explanation on negative mass-square of
   neutrinos, hep.ph/0009291, Fizika B, To appear (2002).
\item Ni, G.-j \& Chang, T.Is neutrino a superluminal particle? Preprint,
hep.ph/0103051.
\item Ni, G.-j, J. Shaanxi Normal Univ. (Natural Science Edition)
29 (3) 1 (2001), hep.th/0201007.
\item Ni, G.-j, Superluminal paradox and neutrino, Preprint,
hep.ph/0203060.
\item Ni, G.-j, Wuli (Physics) 2002, April (4) 255.
\item Ni, G.-j, Evidence for neutrino being likely a superluminal
particle, Preprint, hep-ph/0206.
\item Ni, G.-j. Acta
   Photonica Sinica, 29, 282 (2000); 31, 257 (2002).
\item Ni, G.-j.\& Chen, S.-q. Advanced Quantum Mechanics (Press of Fudan
   University,  March, 2000) chapt.10. The English version of this
   book will be published by Rinton Press in 2002.
\item Ni, G.-j, A new interpretation on quantum mechanics,
quant-th/0206.
\end{enumerate}

\end{document}